\documentclass{aa}  

\usepackage{natbib}
\usepackage{graphicx}
\usepackage[breaklinks=true,colorlinks=true,linkcolor=blue,citecolor=blue]{hyperref}
\usepackage[varg]{txfonts}
\usepackage{amsmath}
\usepackage[usenames,dvipsnames,svgnames,table]{xcolor}
\usepackage{lipsum}
\usepackage{rotating}
\usepackage{txfonts}

\newcommand{\de}{\text{d}}

\newcommand{\msun}{\ensuremath{M_\odot}}
\newcommand{\rsun}{\ensuremath{R_\odot}}

\newcommand{\vel}{{v}}
\newcommand{\lstar}{\ensuremath{L_*}}
\newcommand{\tstar}{\ensuremath{T_*}}
\newcommand{\Teff}{\ensuremath{T_\mathrm{eff}}}
\newcommand{\mdot}{\ensuremath{\dot M}}
\newcommand{\vinfty}{\ensuremath{\vel_\infty}}
\newcommand{\lsun}{\ensuremath{\mathrm{L}_\sun}}
\newcommand{\rstar}{\ensuremath{R_*}}
\newcommand{\mstar}{\ensuremath{M_*}}
\newcommand{\msunyr}{{\ensuremath{\msun}/\mathrm{yr}}}

\newcommand{\kKelvin}{\ensuremath{kK}}

\newcommand{\kms}{\ensuremath{\mathrm{km}/\mathrm{s}}}

\usepackage{float}
\usepackage{hyperref}

\begin{document}

   \title{Polarization of light from fast rotating Wolf-Rayet stars: A Monte Carlo simulations compared to analytical formula}

   \subtitle{}
   \titlerunning{Polarization of light from fast rotating Wolf-Rayet stars}

   \author{S. Abdellaoui
          \inst{1}
          \and
          J.~Krti\v cka
          \inst{1}
          \and
           B. Kub\'atov\'a
          \inst{2}
          \and P.~Kurf\"urst\inst{1}
}
   \institute{Department of Theoretical Physics and Astrophysics,
Faculty of Science,
           Masaryk University, Kotl\'a\v rsk\' a 2, 
Brno, Czech
           Republic,\\ \email{slah@physics.muni.cz}
           \and 
    Astronomical Institute of the Czech Academy of Sciences, Fričova 298, 251 65 Ondřejov, Czech Republic 
           }

   \date{} 
  \abstract
    {Fast-rotating Wolf-Rayet (WR) stars are potential progenitors of long gamma-ray bursts, but observational verification is challenging. Spectral lines from their expanding stellar wind obscure accurate rotational velocity measurements. Intrinsic polarization from wind rotation may help determine rotational speeds, requiring precise wind models.}
    {Our study aims to investigate the intrinsic polarization due to the rotational distortion of  WR winds considering multiple scattering of photons and compare it to a single-scattering model, where we use an analytical expression of the polarization.}
   {We study the polarization signatures resulting from the prolate structure of rotating winds of two WR stars using a 3D Monte Carlo radiative transfer code {\tt Hyperion}. We estimated the intrinsic polarization resulting from multiple scattering in WR winds for different rotational velocities, inclination angles, and mass-loss rates.}
   {Our results indicate that at a rotation rate of less than 50\% of the critical rate, the intrinsic polarization from multiple scattering is close to that of a single scattering model. However, at higher rotation velocities, the polarization from multiple scattering increases with inclination up to 40$^\circ$, while it decreases for inclinations higher than about 60$^\circ$. This dependence is inconsistent with the single-scattering model. We also discuss the effect of the mass-loss rate on the polarization and find that the polarization changes linearly with the mass-loss rate. However, it is important to note that the relationship between polarization and mass-loss rate may vary for different types of stars.}
  {The results have implications for future studies of stellar winds and mass loss and may help to improve our understanding of the complex environments of massive stars.  
 Our research offers valuable information on the complex polarization patterns observed in stellar winds, emphasizing the significance of accounting for the influence of multiple scattering when interpreting observations.}


  \maketitle
%

\section{Introduction}
Classical Wolf-Rayet (WR) stars correspond to a late evolutionary stage of massive stars characterized by the loss of their hydrogen envelope during the evolution \citep{1975MSRSL...9..193C,1979A&A....74...62C,2012A&A...540A.144S}. There are several possibilities for how these stars could lose their envelopes. The classical scenarios include mass-loss by the line-driven winds \citep{2013A&A...558A.131G} and due to binary interactions \citep{2007ApJ...662L.107V}. This would imply that at low metallicities, where line-driven winds become weaker, binary channel dominates, and only the most massive stars reach the domain of WR stars \citep{2024A&A...687A.290G}. However, the absence of binary companions in Small Magellanic Cloud WR stars \citep{2024A&A...689A.157S} indicates the need for an additional mass-loss mechanism that operates in single stars. On the other hand, the WR phenomenon is defined purely on spectroscopic grounds, which may play a role in the apparent independence of binary WR star fraction on metallicity \citep{2020A&A...634A..79S}. 

WR stars have gained significant attention in astrophysics due to their potential role in various cosmic phenomena. Notably, the collapse of rapidly rotating WR stars into black holes has been proposed as a mechanism for generating long-duration gamma-ray bursts \citep{1993ApJ...405..273W}.
Building on this concept, \citet{2005A&A...442..587V} and \citet[among others]{2012A&A...547A..83G} suggested that fast-rotating WR stars could be the progenitors of long-duration gamma-ray bursts. This is supported by radiative transfer models of supernovae explosions associated with gamma-ray bursts, which indicate the breaking of spherical symmetry possibly connected with rotation of the progenitor \citep{2017A&A...603A..51D}. However, directly testing this hypothesis presents challenges, primarily due to the difficulty in accurately measuring the rotational velocities of WR stars. These stars typically show emission lines originating in their expanding envelopes \citep{1995A&A...293..172C,2006A&A...457.1015H}, which prevents a straightforward detection of rotational broadening.

An alternative approach to investigate the rotation of WR stars and thus the origin of gamma-ray burst involves analyzing polarized light from these stars, which may indirectly reveal information about their rotation. As a result of stellar rotation, winds cease to be spherically symmetric with stronger outflows coming from the polar regions due to the effect of nonradial forces and gravity darkening \citep{1996ApJ...472L.115O,2000A&A...358..956P}. The asymmetry can be detected from polarization.
The winds surrounding WR stars are highly ionized, making electron scattering a significant source of opacity in their stellar envelopes. This scattering process transforms the initially unpolarized light from the star's photosphere into linearly polarized light. When light scatters at an angle $\Theta$ relative to its original direction, the ratio of intensities perpendicular and parallel to the scattering plane (defined by the incident and scattered light directions) is $1:\cos^2\Theta$ \citep{1950ratr.book.....C}.
In a perfectly spherically symmetric envelope, the polarization contributions from different directions would cancel each other out, resulting in zero net polarization. However, any asymmetry in the stellar wind or envelope can lead to detectable polarization, potentially providing insights into the star's structure and rotation.

Theoretical modeling of the polarization due to single scattering of a point source radiation illuminating a circumstellar envelope has been studied by \cite{1977A&A....57..141B}.  Their formula was extended to include  the depolarization effect \citep{1987ApJ...317..290C,1989ApJ...344..341B} and stellar occultation \citep{1989ApJ...347..468B}. These models compute the polarization based on the simplified assumption that the photons are scattered only once.  However, it has been shown by \citet{1996ApJ...461..847W,1996ApJ...461..828W}, \citet{2003ApJ...598..572H}, and \citet{2012AIPC.1429..278T} that the polarization obtained from Monte Carlo radiative transfer, which accounts for multiple scattering differs from that obtained using single-scattering model.

In our previous work \citep{2022A&A...658A..46A}, we applied Brown-McLean's formula to compute the polarization from a single-scattering model, assuming optically thin stellar winds from the hydrodynamic calculations. However, in reality, the wind of WR stars is optically thick \citep{2017A&A...608A..34G,2018A&A...614A..86G}. To achieve a more realistic model of the polarization of WR stars, it is necessary to consider the effects of multiple photon scattering. Monte Carlo radiative transfer (MCRT) codes are widely used to model such complex environments because they can account for asymmetric, non-uniform, and optically thick scattering in media. 

In this work, we present a study based on the Monte Carlo (MC) method to investigate the polarization signatures resulting from the stellar wind of two WR stars. We compare the results with the analytical formulation for a single scattering model \citep{1977A&A....57..141B}.

\section{Numerical methods}
\subsection{Wind model setup}

WR stars possess dense and thick winds \citep{2002A&A...389..162N,2017A&A...608A..34G,2018A&A...614A..86G}, making it challenging to use the hydrodynamic model, such as those utilized by \citet{2022A&A...658A..46A}, to accurately predict polarization. These models encounter difficulties in considering the impact of the optically thick wind.

\begin{figure}
    \centering
    \includegraphics[scale=0.6]{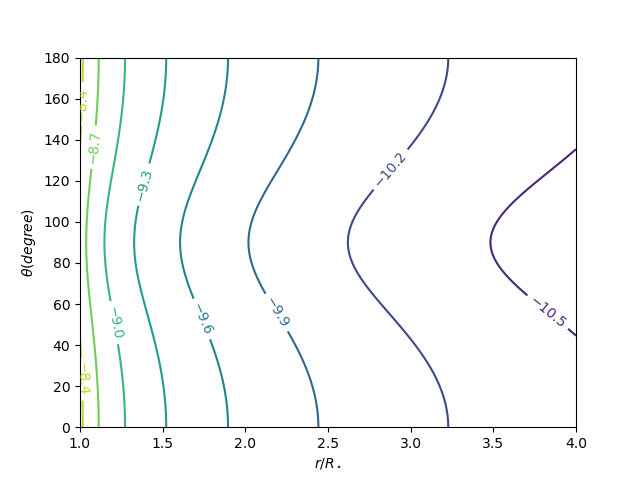}
    \caption{Wind density contours for the rotation rates $\omega=0.7$. The curves are labeled by the logarithm of density in cgs units. }
    \label{fig:dens}
\end{figure}

To address these complexities, we use an alternative model developed by \citet{2002ApJ...581.1337D}. This model is specifically tailored to tackle the extreme conditions characteristic of WR stars, accounting for both gravity darkening and the influence of optically thick winds. Consequently, this model is better suited for our polarization calculations.
The mass flux is considered to be higher at polar regions decreasing toward the equator,  due to the gravity-darkening effect. Using the model of \citet{2002ApJ...581.1337D}, the stellar wind velocity is expressed as 
\begin{equation}
    \vel(r,\theta)= \vel(r,\theta=0)\sqrt{1-\omega^2\sin^2\theta},
\end{equation}
and the mass flux $\dot{m}(\theta)$ as
\begin{equation}
    \dot{m}(\theta)= \dot{m}(1-\omega^2\sin^2\theta).
\end{equation}
The wind density distribution (see Fig.~\ref{fig:dens}) is given by 
\begin{equation}
    \rho(\omega,\theta)=\rho(r,\theta=0)\sqrt{1-\omega^2\sin^2\theta},
\end{equation}
where $\rho(r,\theta=0)=\dot{M}/(4\pi v(r,\theta=0) r^2)$, $\dot{M}$ is the wind mass-loss rate, $\omega=\Omega/(\rstar v_\mathrm{crit})$ is the rotation rate (the ratio of angular velocity $\Omega$ to the critical velocity), $r$ is the radial distance in the spherical coordinate, $\theta$ is co-latitude, and $v(r,\theta=0)$ is the so-called $\beta$-velocity law \citep[see e.g.][]{Lamers:1999} described as $v(r,\theta=0)=v_0 + v_\infty(1-\rstar/r)^{\beta}$, where $v_0=0.1v_\infty$ is the initial wind velocity, $v_\infty$ is the terminal wind velocity, the exponent $\beta$ is a parameter describing the steepness of the velocity law fixed to the standard value $\beta=1$, and $\rstar$ is the stellar radius. For the region located between the point source and the base of the wind at $r=\rstar$, we assumed a zero density.
 The critical angular velocity is calculated using
\begin{equation}
    v_\mathrm{crit}=\sqrt{\frac{G\mstar}{\rstar}}, 
\end{equation}
with G being the gravitational constant.
Table~\ref{tab:star} shows the stellar and wind parameters of two WR stars used in this work.

\begin{table}
\caption{Adopted stellar and wind parameters \citep{2015A&A...581A.110T,2018MNRAS.479.4535S}.}
\label{tab:star}
\begin{tabular}{ccccccc}
\hline 
\hline 
WR & $\log{\lstar/\lsun}$ & $\tstar$ & $\rstar$ & $\mstar$ & $\vinfty$ & $\log{\mdot}$ \\ 
&&    ($\kKelvin$) & ($\rsun$) & ($\msun$) & ($\kms$) & ($\msunyr$) \tabularnewline
\hline 
93b & 5.30 & 160 & 0.58 & 
7.1 & 5000 & -5.00 \tabularnewline
102 & 5.45 & 210 & 0.39 & 
7.0 & 5000 & -4.92\tabularnewline
\hline 
\end{tabular}
\end{table}
\subsection{Continuum polarization calculation}

\subsubsection{Analytical calculations}
To compute the polarization analytically in the winds of WR stars we use the mathematical expression derived by \cite{1977A&A....57..141B} as
\begin{equation}
    \label{polarcor}
        P _R= \frac{3}{16} \sigma_\text{T}\sin^2 i \int_{\rstar}^{r_f}\int_{-1}^{1}n_e(r,\mu)(1-3\mu^2)D(r) \, \de r \, \de\mu, 
\end{equation}
where $\sigma_\text{T}$ is the Thomson scattering cross section, $\mu$ is the
cosine of the polar angle, $n_e(r, \mu)$ is the electron number density of the envelope, and $i$ is the inclination angle of the symmetry axis with respect to the observer 
\citep[cf. also the calculation of polarization
signatures in][]{2020A&A...642A.214K}. 
$D(r)$ is the depolarization factor introduced by \cite{1987ApJ...317..290C} as $D(r)=\sqrt{1-\rstar^{2}/r}$. The lower limit of integration is $r=\rstar$ and the upper limit corresponds to the outer boundary of numerical simulations, $r_f=10~\rstar$.
                                               
Since the wind of WR stars is optically thick to electron scattering, we include the attenuation factor $e^{-\tau}$ \citep{1979MNRAS.186..265M,1986ApJ...303..292F}, where optical depth $\tau$ is given as
\begin{equation}
    \tau = \sigma_T \int_{\rstar}^{r_f}n_e(r,\mu)\de r.
\end{equation}
The analytical expression of polarization becomes
\begin{equation}
    \label{pr}
        P _R= \frac{3}{16} \sigma_\text{T}\sin^2 i \int_{\rstar}^{r_f}\int_{-1}^{1}n_e(r,\mu)(1-3\mu^2)D(r)e^{-\tau} \, \de r \, \de\mu. 
\end{equation}

\subsubsection{Monte Carlo calculations} 
The continuum polarization in hot star winds appears due to the scattering of photons by free electrons in an axisymmetric environment. The degree of polarization depends on several parameters, such as the optical depth, and the inclination. The polarization state of radiation can be described by the Stokes vector $S$ \citep{1950ratr.book.....C} as follows, 
\begin{equation}
S=\left(\begin{array}{c}
I\\
Q\\
\begin{array}{c}
U\\
V
\end{array}
\end{array}\right),
\end{equation}
where $I$ is the radiation intensity, $Q$ and $U$ represent the linear polarization and $V$ describes the circular polarization. The degree of linear polarization can be written as
\begin{align}
P_R & =\sqrt{Q^{2}+U^{2}}/I,
\end{align}
and the angle of linear polarization is given by
\begin{equation}
    \tan\psi=\frac{1}{2}\frac{Q}{U}.
\end{equation}
To record the change in polarization state during a scattering event, the Stokes vector is multiplied by the M\"uller matrix, $M$, corresponding to the event. It is assumed that the reference direction lies in the scattering plane and the plane orthogonal to the propagation direction. The components of the M\"uller matrix depend on the geometry of the scattering event, the physical properties of the scatterer, and often on the wavelength. The M\"uller matrix  for rotation is written as \citep{1989fsa..book.....C}
\begin{equation}
M(\phi)=\left(
\begin{array}{cccc}
1 & 0 & 0 & 0\\
0 & \cos2\phi & -\sin2\phi & 0\\
0 & \sin2\phi & \cos2\phi & 0\\
0 & 0 & 0 & 1
\end{array}
\right).
\end{equation}
For electron scattering, M\"uller matrix can be expressed as a function of the scattering angle $\Theta$ \citep{1950ratr.book.....C,1995ApJ...441..400C,2017A&A...601A..92P}
\begin{equation}
M_\text{Th}(\Theta)=\frac{3}{4}\left(
\begin{array}{cccc}
\cos^{2}\Theta+1 & \cos^{2}\Theta-1 & 0 & 0\\
\cos^{2}\Theta-1 & \cos^{2}\Theta+1 & 0 & 0\\
0 & 0 & 2\cos^{2}\Theta & 0\\
0 & 0 & 0 & 2\cos^{2}\Theta
\end{array}    \right).
\end{equation}
We use a 3D MCRT code {\tt Hyperion}\footnote{{\tt Hyperion} is published under an open-source license at http://www.hyperion-rt.org.} \citep{2011A&A...536A..79R} to calculate the polarization of two fast rotating WR stars (WR 93b and WR 102, see Table~\ref{tab:star}), considering multiple scattering.
 The code is developed to cover a wide range of problems. It is parallelized and solves the radiative transfer equation in various geometries, including Cartesian, cylindrical, polar, spherical, and adaptive Cartesian grids. The code also computes the temperature, spectral energy distribution, and images. The MC method solves the radiative transfer equation by simulating photon packages and using a ray tracing approach.

 {\tt Hyperion} code uses a 4-element M\"uller matrix to calculate the polarization taking into account multiple scattering. After the photon has left the computational domain, the reference direction of the Stokes vector is adjusted to align with the star's rotational axis. This alignment allows us to group the photons into specific observer bins corresponding to the appropriate inclination. A total of $10^7$ photons were utilized for imaging and ray tracing.
 For a direct comparison between our Monte Carlo simulations and the analytical formula by \citet{1977A&A....57..141B}, the stars are represented as point sources of known radius ($\rstar$), luminosity ($\lstar$), and effective temperature ($\Teff$), located at the center of the computational domain (i.e. the winds of the WR stars) with the lower boundary at $r=\rstar$ and the outer boundary at ${r_f}=10~\rstar$.
 This assumption separates the effects of scattering in the stellar wind, which is the primary focus of our work, and ensures that any differences between the simulations and the analytical outcomes can be linked to the scattering physics rather than the geometry of the source. Although a spherical source would give a more accurate representation of the star, it brings a more complexities (e.g., limb darkening, gravity darkening) that are unnecessary for the study in this paper.\\
 
\section{Rotational effect} 
\begin{figure}
    \centering
   \includegraphics[scale=0.6]{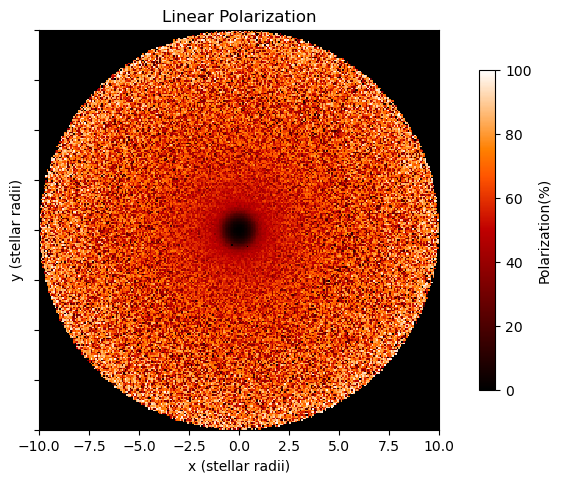}
    \caption{Polarization 
    map as a function of stellar radius for WR93b star with rotation rate $\omega=0.6.$}
    \label{fig:maps}
\end{figure}

\begin{figure*}
     \includegraphics[width=0.5\textwidth]{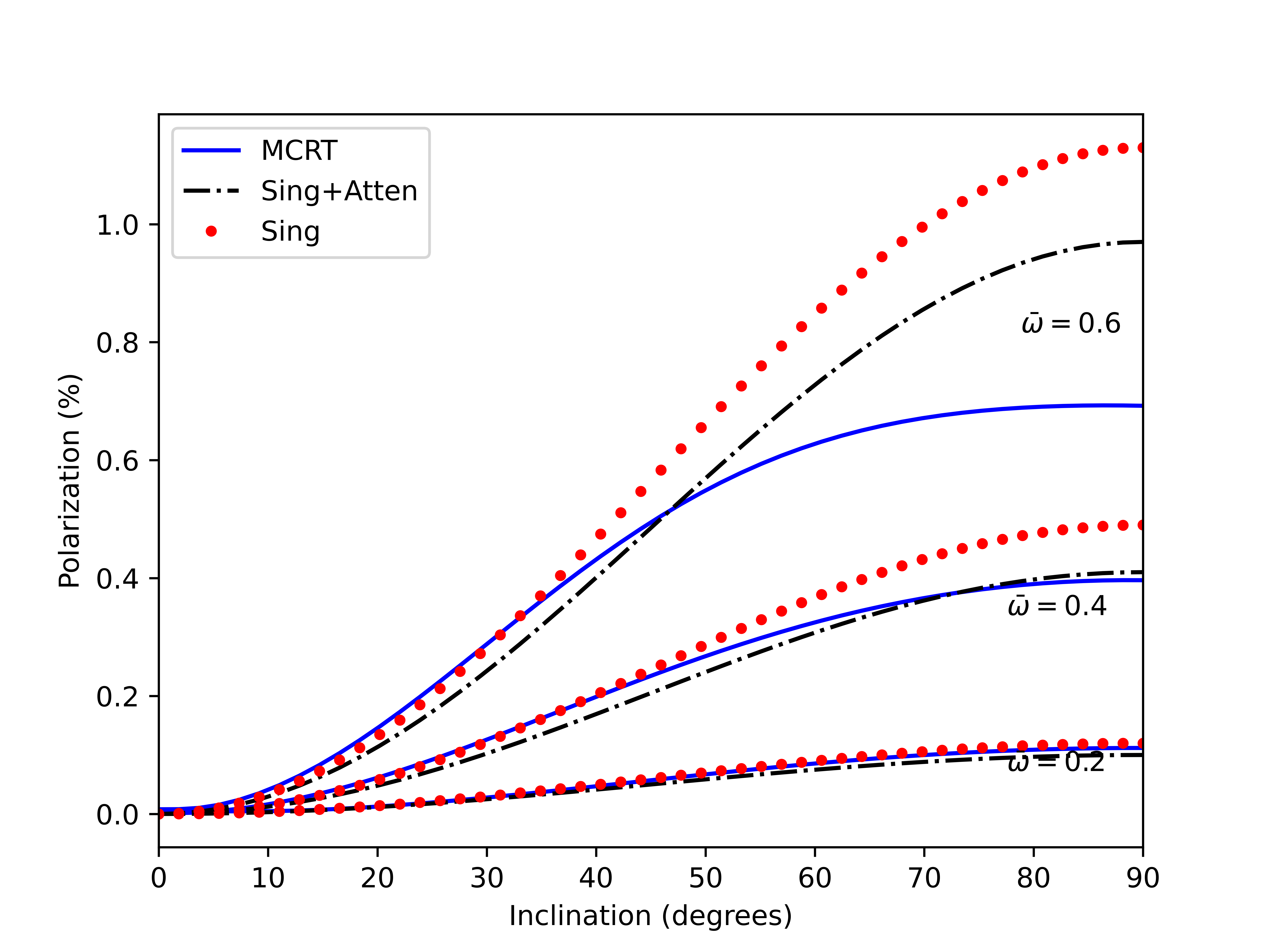}
\includegraphics[width=0.5\textwidth]{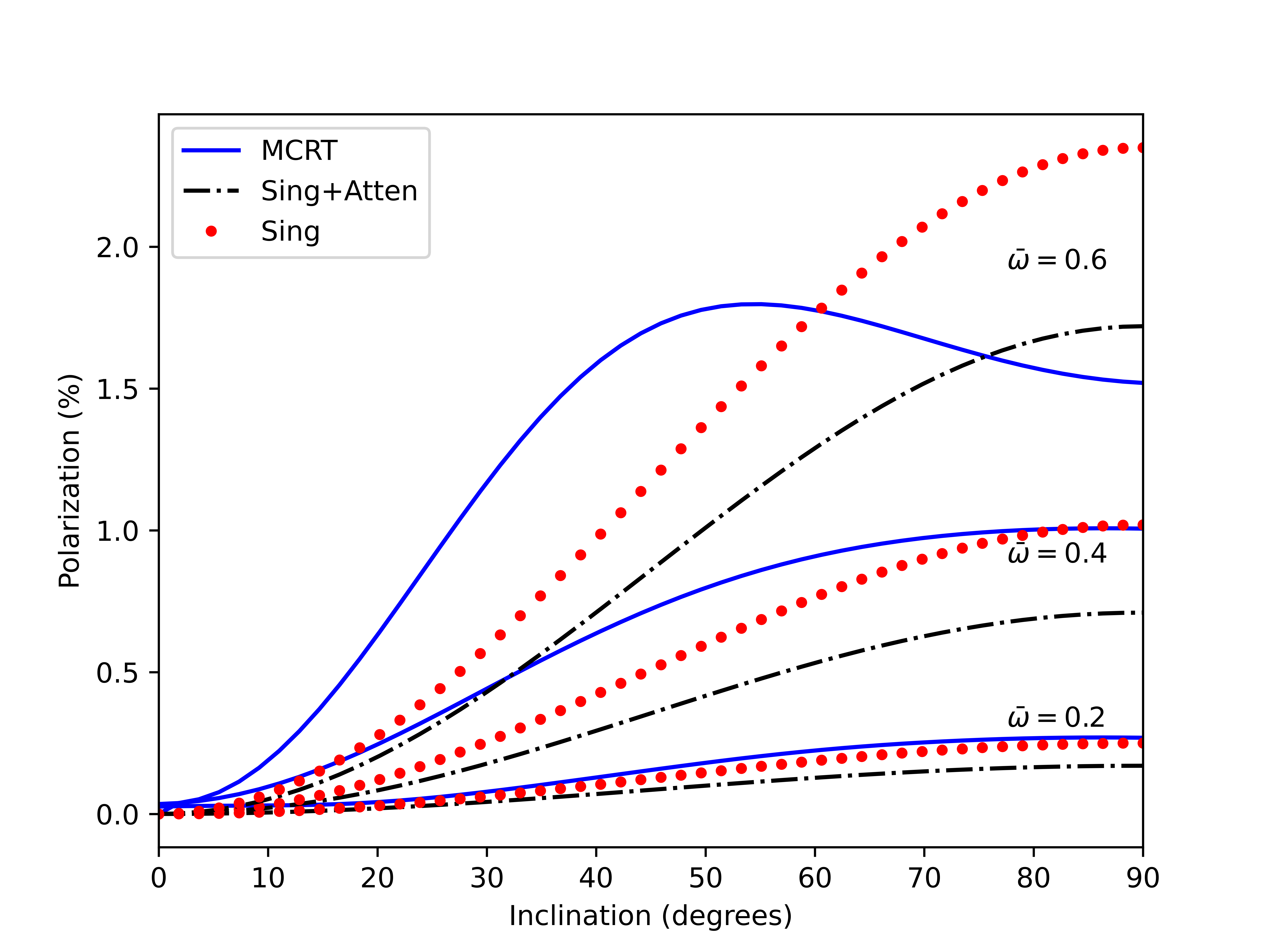}   
    \caption{Polarization as a function of inclination for single-scattering model (red dotted lines), single-scattering model with attenuation (black dashed-dotted lines), and multiple scattering model (full blue lines) of stars WR93b (left panel) and WR102 (right panel) computed for a fixed mass-loss rate as it is given in Table~\ref{tab:star}.}
    \label{fig:compsingmul}
\end{figure*}

In this section, we demonstrate the effect of rotation on the polarization of scattered light. We conduct simulations for various values of $\omega$. The obtained polarization distribution in the ($x, y$) plane by MCRT simulation is shown in Fig.~\ref{fig:maps}.
The results from comparing the single-scattering model (Eq.~\ref{polarcor}), the single-scattering model with attenuation (Eq.~\ref{pr}), and the multiple-scattering model with MCRT code are displayed in Fig.~\ref{fig:compsingmul}. These results show that the three models agree reasonably well up to an angular velocity of 50 \%  of the critical value. However, as the rotation increases beyond this point, there are noticeable deviations in the polarization induced by multiple scattering compared to the single-scattering model.
 
\begin{figure*}
    \centering
        \includegraphics[width=0.49\textwidth]{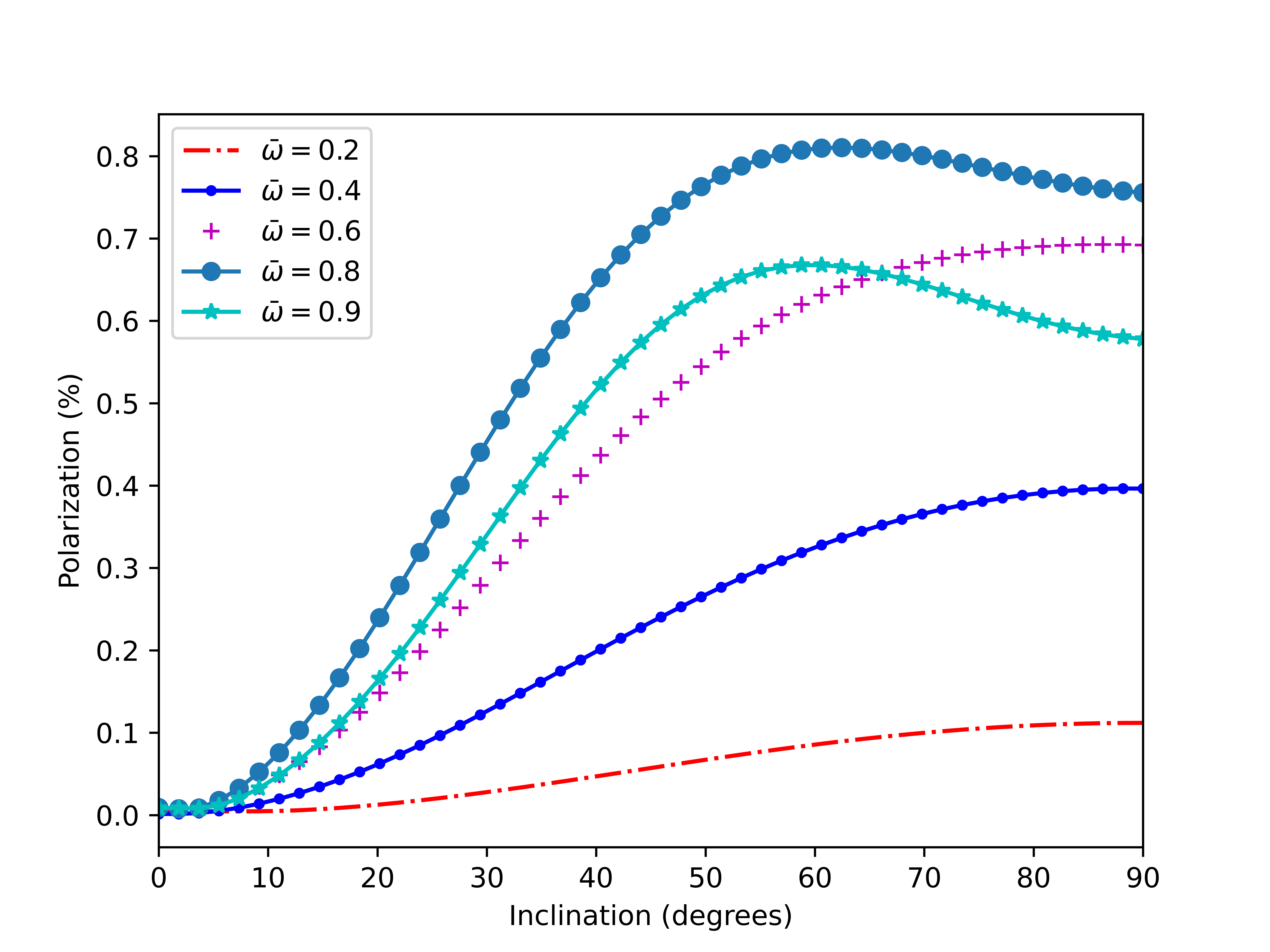}
        \includegraphics[width=0.49\textwidth]{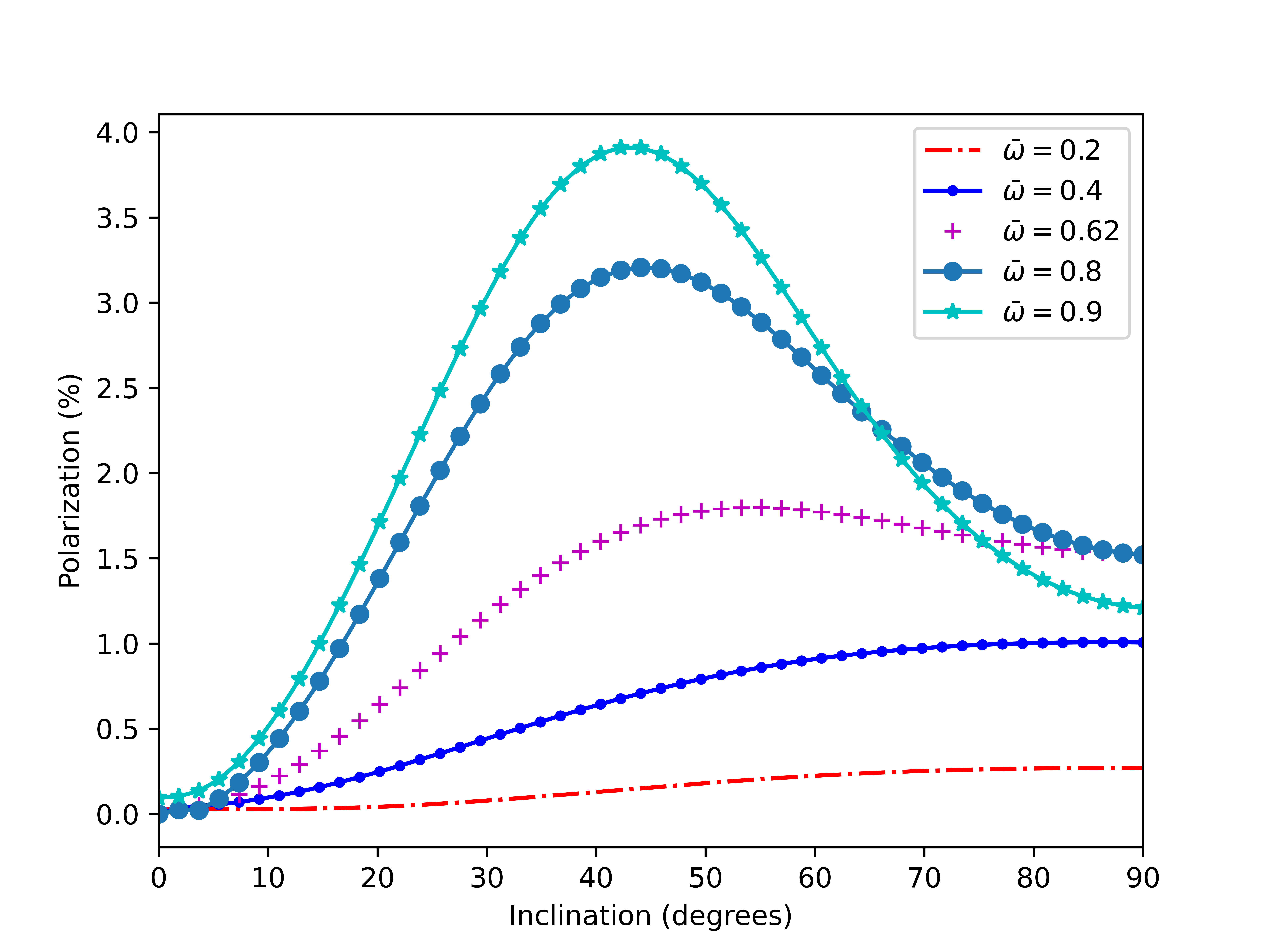}
    \caption{Polarization as a function of inclination for the multiple-scattering model of stars WR93b (left panel) and WR102 (right panel) with different rotation rates, as labeled in the panels.}
    \label{fig:mcrt}
\end{figure*}

Upon closer examination of the polarization data, it becomes evident that the impact of multiple scattering is particularly pronounced at inclination angles of 40 and 60$^\circ$. The degree of polarization experiences a significant increase at 40$^\circ$, followed by a sharp decrease at 60$^\circ$, as shown in Fig.~\ref{fig:mcrt}. 
Similar results were obtained by \citet{1996ApJ...461..828W, Halonen_2013}, where the polarization peaks at an inclination lower than 90$^\circ$. 

\begin{figure*}
    \centering
    \includegraphics[width=0.49\linewidth]{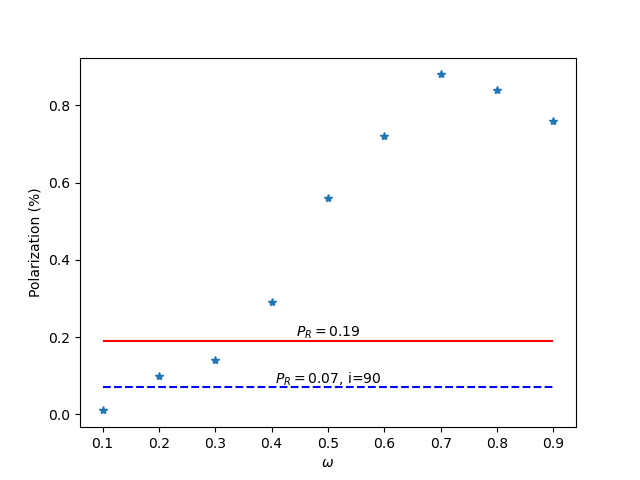}
    \includegraphics[width=.49\linewidth]{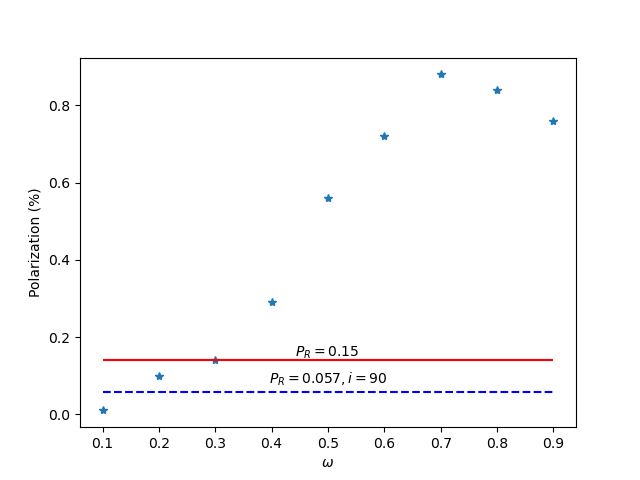}
    \caption{Polarization as a function of rotation $\omega$ from the multiple-scattering model viewed edge-on  for WR93b (left panel) and WR102 (right panel). The blue dashed lines represent the observed polarization at an inclination $i=90^{\circ}$ and the red lines represent the scaled polarization by $<\sin^2 i>$ from \citet{2018MNRAS.479.4535S}. }
    \label{fig:polrot}
\end{figure*}

Figure~\ref{fig:polrot} illustrates the effect of rotational velocity on polarization for an edge-on view. The graph demonstrates an increase in polarization as the rotation rate $\omega$ increases, reaching a peak around $\omega=0.7$, similar to the result in Fig.~\ref{fig:mcrt}.
Moreover, the observed upper limit of polarization \citep{2018MNRAS.479.4535S} for both WR93b ($P_R<0.077$) and WR102 ($P_R<0.057$) indicates a rotation rate $\omega < 0.2$, corresponding to rotational velocities of less than 347\,\kms\ and 457\,\kms, respectively. By using the formula $j=\varv_\text{rot}\rstar$ to calculate the specific angular momentum, we find that $\log(j/(1\mathrm{cm}^2/\mathrm{s})<18$. We note that these values of angular momentum are comparable to the one obtained by \citet{2018MNRAS.479.4535S}, and \citet{2022A&A...658A..46A}. For the scaled polarization by $<\sin^2 i>$, we can observe that the relative rotation rates are $\omega\approx0.25$, and $\omega\approx0.3$, which correspond to a rotational velocity of $\varv_\text{rot}\approx433\,\kms$, and $\varv_\text{rot}\approx 686\,\kms$, for WR93b and WR102, respectively. 

According to the threshold set by \citet{1999ApJ...524..262M}, for a star to collapse into a LGRB, the angular momentum $j$ must exceed $3\times10^{16}\,\mathrm{cm}^2/\mathrm{s}$. As our calculated values exceed this threshold, it can be inferred that these WR stars are potential candidates for LGRB. 
\section{Mass loss effect}
Various factors influence the polarization properties of WR stars, and one important factor is the mass-loss rate. The mass-loss rate of a WR star determines the optical depth of its stellar wind, which affects the polarization of the scattered light.

\begin{figure*}
    \centering
\begin{minipage}{0.48\linewidth}
        \includegraphics[scale=0.5]{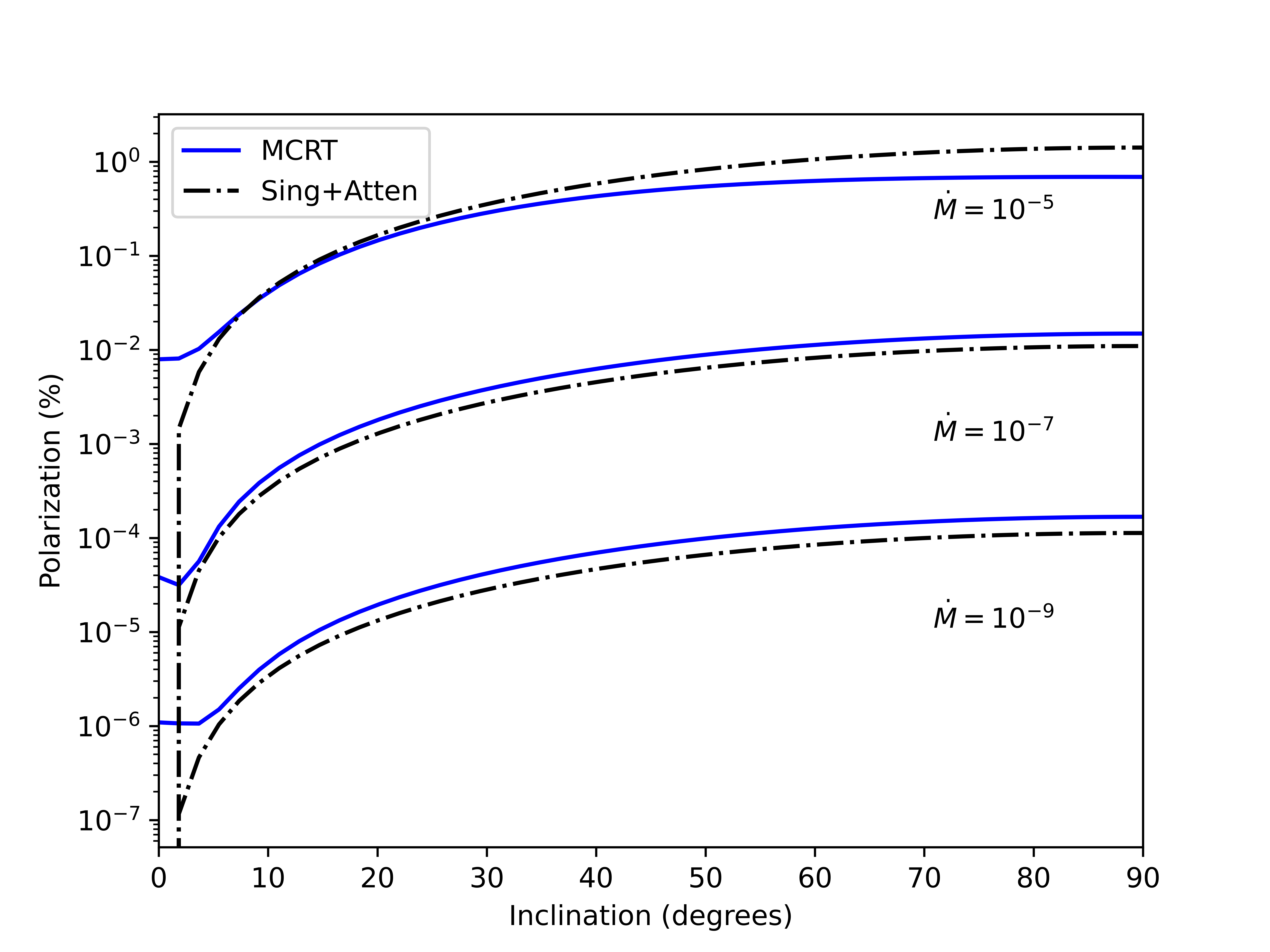}
\end{minipage}
\begin{minipage}{0.48\linewidth}
    \includegraphics[scale=0.5]{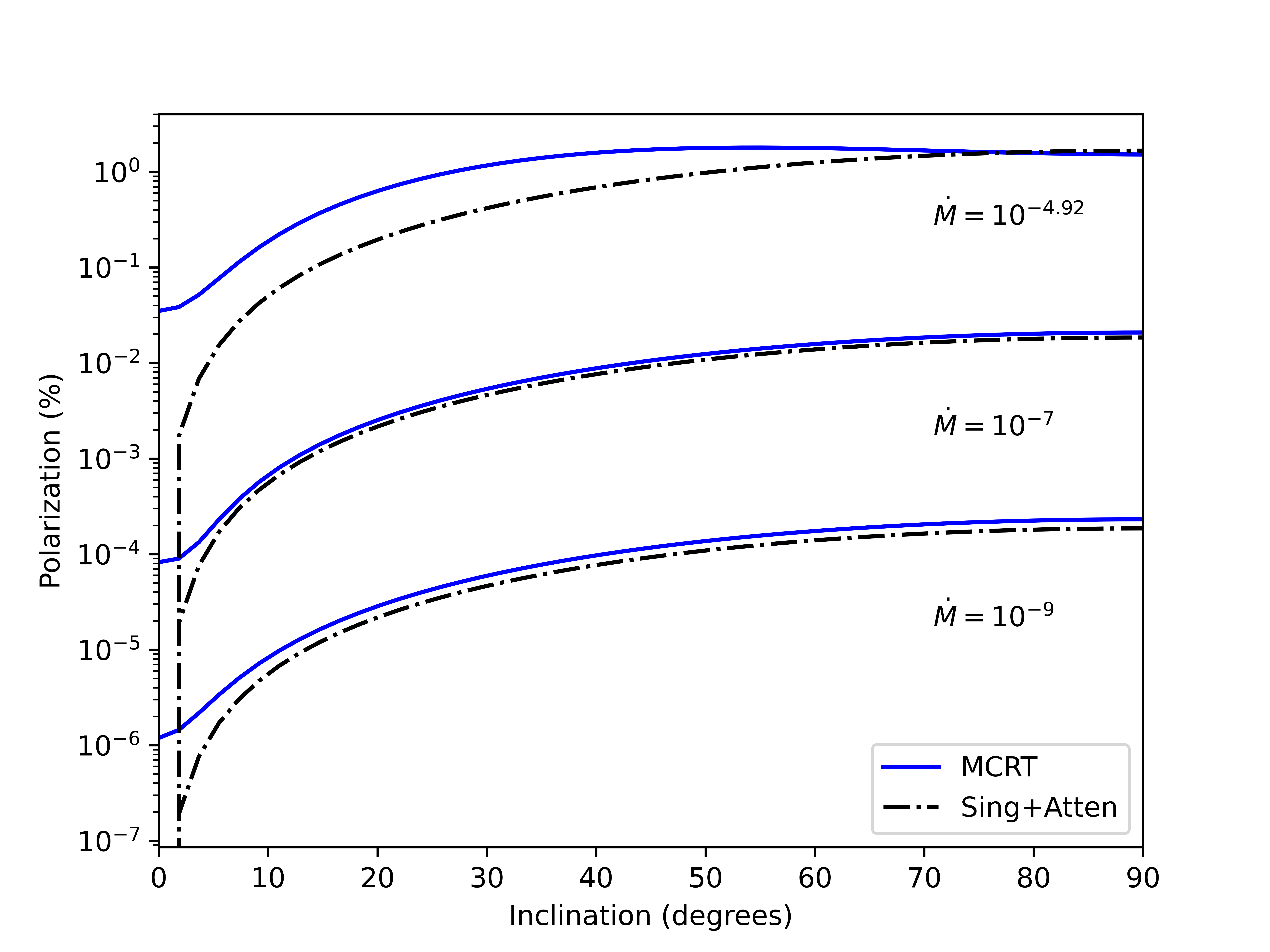}
\end{minipage}
    \caption{Comparison of polarization from the multiple-scattering model (MCRT simulation, full blue lines) and single scattering (Sing+Atten) model with attenuation (black dashed-dotted lines) as a function of inclination for WR93 (left panel) and WR102 (right panel). The blue and black lines are labeled with the mass-loss rates in the units of $M_\odot/\text{yr}$.}
    \label{fig:mdot}
\end{figure*}

To better understand this relationship, we examine a plot that shows the polarization as a function of inclination for different values of mass-loss rate. 
We found that as the mass-loss rate increases, the degree of polarization also increases (see Fig.~\ref{fig:mdot}). This phenomenon occurs because higher mass-loss rates result in more scattering particles in the stellar wind, leading to higher polarization.

Therefore, the mass-loss rate plays a significant role in determining the polarization properties of WR stars. By understanding and studying this relationship, we can gain valuable insights into the physical processes occurring in these stars and further our understanding of their evolution and characteristics.

\section{Conclusion}
Our study examined the polarization resulting from electron scattering in a rotating stellar wind. We looked at cases where a point source of radiation illuminated the wind. We study how different parameters affected the polarization behavior when there was only scattering or scattering with absorption.

We found that the polarization strongly depends on the viewing angle for all models. Interestingly, our simulations revealed that multiple scattering had a significant impact on the polarization compared to what would be predicted by analytical models assuming single scattering. 

For cases with shallow optical depths, our simulations were consistent with the $\sin^2 i$ dependence of polarization described by \citet{1977A&A....57..141B}. However, we observed that many of our models exhibited a peak at lower inclination angles for high optical depth. 

Our study provides insights into the complex polarization behavior in stellar winds and highlights the importance of considering multiple scattering effects when interpreting observations.

The obtained polarization is comparable to previous calculations, and the limit for a star to potentially produce a long-duration gamma-ray burst (LGRB) is satisfied, with the angular momentum that could be greater than $3\times10^{16}\,\mathrm{cm}^2/\mathrm{s}$. These findings suggest that WR stars like WR93b and WR102 could indeed be the progenitors of LGRBs. 
Therefore, external factors such as magnetic fields, interactions with binary systems, or mass loss may significantly impact their ultimate fate. Further research involving spectropolarimetry and sophisticated modeling is crucial to better understand the role of these stars in LGRB progenitor populations.

In this study, we applied the density model formulated by \citet{2002ApJ...581.1337D}, which originates from the gravity darkening theorem introduced by \citet{vonZeipel1924}. A more accurate description of gravity darkening has been developed for fast rotating stars by \citet{2011A&A...533A..43E}; the von Zeipel model remains a practical and widely used approximation for exploring the effects of rotation on stellar winds.
For stars with rapid rotation, the more accurate description results in a reduced temperature ratio between poles and the equator, consequently causing a lower ratio of mass-loss rates in these regions. Such lower mass-loss rare ratio leads to a slightly lower polarization than predicted here.
\begin{acknowledgements}
The authors thank Prof.~R.~Ignace for the useful discussion.
We gratefully acknowledge support from the Grant Agency of the Czech Republic (GAČR 25-15910S). The Astronomical Institute of the Czech Academy of Sciences in Ond\v rejov is supported by the project RVO:67985815.
Computational resources were provided by the e-INFRA CZ project (ID:90254),
supported by the Ministry of Education, Youth and Sports of the Czech Republic.
\end{acknowledgements}
\bibliographystyle{aa} 
\bibliography{references}
\label{LastPage}
\end{document}